# The Joint astrophysical Plasmadynamic EXperiment (J-PEX): A high-resolution rocket spectrometer

M.A. Barstow[*a], N.P. Bannister[a], R.G. Cruddace[b], M.P. Kowalski[b], K.S. Wood[b], D.J. Yentis[b], H. Gursky[b], T.W. Barbee, Jr.[c], W.H. Goldstein[c], J.F. Kordas[c], G.G. Fritz[d], J.L. Culhane[e], and J.S. Lapington[e,f]

[a]University of Leicester; [b]Naval Research Laboratory; [c]Lawrence Livermore National Laboratory; [d]PRAXIS Inc.; [e]Mullard Space Science Laboratory; [f]Boston University

**ABSTRACT**

We report on the successful sounding rocket flight of the high resolution (R=3000-4000) J-PEX EUV spectrometer. J-PEX is a novel normal incidence instrument, which combines the focusing and dispersive elements of the spectrometer into a single optical element, a multilayer-coated grating. The high spectral resolution achieved has had to be matched by unprecedented high spatial resolution in the imaging microchannel plate detector used to record the data. We illustrate the performance of the complete instrument through an analysis of the 220-245Å spectrum of the white dwarf G191-B2B obtained with a 300 second exposure. The high resolution allows us to detect a low-density ionized helium component along the line of sight to the star and individual absorption lines from heavier elements in the photosphere.

## 1. INTRODUCTION

Beginning with sounding rocket-borne experiments in the late 1960s and early 1970s, observations in the Extreme Ultraviolet (EUV) wavelength range (spanning ~100-1000Å) have proved to be of great importance for astrophysics. Development of the field of EUV astronomy has been slow, due to a combination of technical challenges with the necessary instrumentation and a belief, subsequently proved to be erroneous, that the interstellar medium was completely opaque to EUV photons. However, with the detection of the first known EUV sources by the Apollo Soyuz Test Project and launch of the Einstein and EXOSAT X-ray observatories, which had a capability overlapping into the EUV, the usefulness of the waveband became apparent. All sky surveys for EUV sources were carried out by the ROSAT Wide Field Camera (WFC, 1990) and the Extreme Ultraviolet Explorer (EUVE, 1992), with the latter mission providing spectroscopic follow-up until it was shut down in February 2001. A detailed history of the development of EUV astronomy is now available[1].

Approximately 1000 sources of EUV radiation are now known and moderate resolution (R~300) spectroscopy has been performed on the brightest of these. However, despite the tremendous success of EUVE, its limited sensitivity and spectral resolution have prevented exploration of some important scientific topics. For example, in those hot hydrogen-rich DA white dwarfs containing significant quantities of trace heavy elements, it has not been possible to resolve the large numbers of absorption lines present. This has prevented a complete identification of all the possible absorbing species. More importantly, these heavy element lines mask the possible signature of a contribution from interstellar and/or photospheric ionized helium. The presence of such a component is inferred from EUVE spectroscopy of several stars[2,3], but is not directly detected. It was also difficult to carry out time-resolved spectroscopic observations with EUVE, since the exposure times required to achieve an appropriate signal-to-noise were often longer than the timescales of the phenomena being studied.

New advances in the field of EUV astronomy require access to instrumentation of considerably more sensitivity and spectral resolution than has hitherto been available. The Joint astrophysical Plasmadynamic Experiment (J-PEX) is a sounding rocket-borne spectrometer designed to answer this need. Based on normal-incidence optics, it is hoped that the instrument will serve as the proto-type for a new generation orbiting observatory. We report here on the instrument design and the results from a first successful flight, carried out in February 2001.

## 2. THE J-PEX SPECTROMETER DESIGN

**2.1 Optical layout and structure**

The spectrometer is a slitless, normal-incidence instrument, which employs a figured spherical grating in a Wadsworth mount. As a result of practical upper limits on the grating size at the time the payload was constructed, four individual identical grating segments were utilized to maximize the geometric collecting area. The design of the J-PEX payload is shown in Fig. 1. The EUV light from the star enters a collimator, which minimizes the EUV background flux into the spectrometer, and strikes the grating at an angle of incidence of 4.85 deg. The grating is coated with a multilayer designed for high efficiency in the band 220-245 Å. The diffracted radiation is focused onto an MCP detector mounted on the grating optical axis, a configuration, which minimizes aberrations. Diffracted and scattered EUV/FUV background is trapped by baffles (not shown) within the spectrometer, while residual background reaching the detector is attenuated by an aluminum filter. Two telescopes measure the small target motions due to drift and jitter of the attitude control system (ACS). One has a CCD camera at the focus and operates at optical and UV wavelengths. The other has a multilayer-coated mirror, which focuses an EUV image of the target onto the MCP detector.

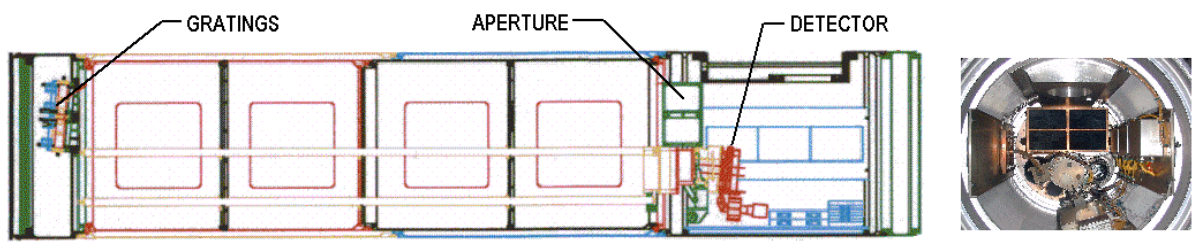

Fig. 1: The J-PEX high-resolution spectrometer payload, designed and constructed for launch by a Terrier boosted Black Brant IX sounding rocket.

The structural design primarily comprises two concentric shells, a forward end closure and a motorized door covering the aperture, which is opened following separation from the Black Brant sustainer motor. The payload is pumped on the launch pad prior to launch to ensure that the MCP detector can operate in a vacuum environment ($\sim 10^{-6}$ mbar) after its own vacuum door opens in flight. The 22-inch outer shell is standard Black Brant structure, which mates to the NASA payload systems. The inner shell forms the optical bench, upon which the optics and the detectors are mounted. The two joints between the shells incorporate compressed O-rings, providing dynamic and thermal isolation of the inner from the outer shell, and reduce bending loads. The grating and the optical mirror are mounted together rigidly at one end of the bench. The MCP and CCD detectors are likewise mounted together rigidly, but on a structure, which is connected to the optics plate by Invar tubes. In the presence of thermal expansion and slight flexing introduced by temperature changes, this design ensures stable co-alignment of the spectrometer and telescopes and maintains a fixed focal length.

**2.2 Gratings**

The large collecting area required for astrophysical observations is achieved by using four grating segments, each 9 cm x 16 cm x 2cm, prepared by Zeiss. Each blank of Herasil fused silica was ground to produce a spherical surface of radius 4 m (2 m focal length), followed by superpolishing to yield a surface microroughness of 3.3 – 4.3 Å (RMS). Ion-etching produced grooves having a density of 3600 g mm$^{-1}$ and a depth (62 Å) designed to suppress zero order and maximize 1$^{st}$ order efficiency at 235 Å. The measured groove efficiency ($\eta_{gr}$) for two gratings was 0.34, close to the maximum theoretical value 0.405. For the other two $\eta_{gr}$ was degraded inadvertently, following a special optics cleaning process used routinely before multilayer coating. This problem has since been overcome, but reduced the average J-PEX groove efficiency to 0.285 (Table 1). The multilayer coatings ($Mo_5C/Si/MoSi_2$) were prepared by Dr. Troy Barbee of the Lawrence Livermore National Laboratory and achieved a maximum reflection efficiency of 0.30 at the center of the 225-245 Å passband, with a final surface microroughness of 3.3 Å. The total efficiency (groove x multilayer) of one of the best two gratings is shown in Fig. 2, as function of wavelength, and the overall performance for the group of 4 summarized in Table 1.[4,5]

| Parameter | Value |
|---|---|
| Grating aperture area ($cm^2$) | 456 |
| Grating groove efficiency | 0.285 |
| Multilayer peak efficiency | 0.285 |
| MCP photocathode efficiency | 0.15 |
| Collimator transmission | 0.90 |
| EUV/FUV filter transmission | 0.50 |
| Effective area ($cm^2$) | 2.5 |

Table. 1: Grating aperture area, subsystem efficiencies and transmission coefficients for the J-PEX spectrometer.

**2.3 MCP detector**

The MCP focal plane detector is mounted in its own vacuum chamber equipped with an ion pump, and has a motor-operated vacuum door developed for the NRL Special Sensor Ultraviolet Limb Imager (SSULI). Front-end imaging electronics and the MCP high-voltage supply are mounted on the back of the detector. The original design aimed at a significant advance in spatial resolution using two square 35 x 35 mm Photonis MCPs in a chevron configuration. The pores had an internal diameter of 6μm and a length-to-diameter ratio of 120, and the front MCP was coated with a cesium iodide (CsI) photocathode. Charge pulses are detected by a Vernier anode located behind the MCP stack. This readout employs a repeated sequence of nine linear anodes deposited on a multilayered substrate. The area of each anode varies along its length in a cyclic manner, with wavelength and phase varying among the anodes. Analysis of the charge collected by the anodes yields a two-dimensional image.

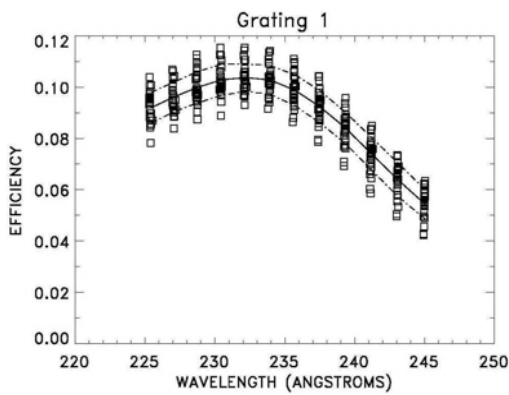
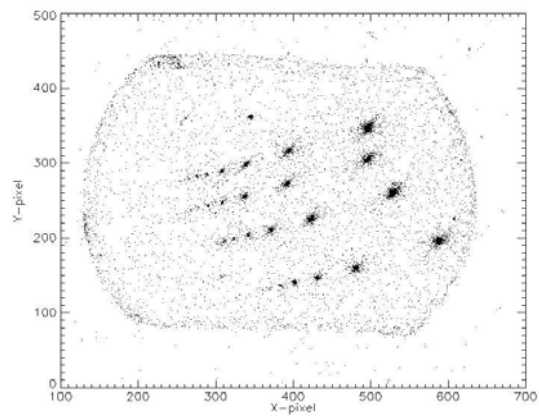

Fig. 2: The efficiency of a multilayer-coated J-PEX flight grating in first inside order, measured as a function of wavelength at 29 points on the grating. The efficiency is close to the maximum allowed theoretically.

Fig. 3: Spectra obtained during the focusing of the four J-PEX gratings, showing the emission lines of ionized helium. The strongest line visible is at 243.03 Å. The images have an intense saturated core with a weak halo.

The detector underwent a series of tests, including functional and vibration tests, measurement of image linearity and spatial resolution, and calibration of the quantum efficiency (QE). However a major difficulty arose in achieving a satisfactory QE using the small-pore MCPs. This has been a widespread but unpredictable characteristic of MCPs working in the EUV, caused by an insufficient photoelectron yield in the MCP pores. Recent reports suggest that high QE (0.35-0.40) has been achieved in the EUV by careful control of the MCP preparation, but these results have proved difficult to reproduce by adjusting the processing of the small-pore MCPs. A pragmatic solution to the problem was achieved, at the expense of spatial resolution, by replacing the front MCP with a 12μm pore, 25mm diameter circular plate made at the IKI in Russia, and coated with a CsI photocathode. EUV calibrations yielded a QE of 14.6% in the 220-245 Å band.

Image uniformity and linearity were examined using a pinhole array (15μm diameter, spacing 500μm) placed close to the front MCP. These data were used to optimize the real-time position-decoding algorithm for minimum non-linearity. Typically the Vernier anode in charge collection mode exhibits non-linearities < 20μm RMS. Lack of a collimated beam constrained measurements of spatial resolution, and while they suggested a value approaching 15μm (FWHM) was achieved, in the data analysis we assume a firm upper limit of 20μm.

**2.4 Collimator and background filter**

The background in the focal plane detector has three main causes, namely the intrinsic MCP background, cosmic rays and UV background. The latter is the strongest, and using the review by Meier[6] and an analysis of EUVE experience by Drake (1997), we found the dominant lines to be H I (1215.7 Å), O II (833.8 Å), He I (584.3 Å), He II (303.8 Å) and O II (303.7 Å). UV background flux entering the instrument is reduced by a honeycomb collimator installed in the aperture. It was fabricated from copper foil[7], using techniques developed for the NRL X-ray detectors on the DoD ARGOS satellite. The collimator has an approximately circular field-of-view 1.15 deg in diameter (FWHM), and its surfaces have been blackened with Ebonol-C, which has low reflectivity in the far UV. Most of the background admitted by the collimator, which is diffracted and scattered by the gratings, is intercepted by light baffles. The background radiation entering the detector aperture is then primarily residual scattered light, which is attenuated by an aluminum filter. The thin film filter is supported on a nickel mesh, and is under vacuum during launch and therefore suffers no acoustic loads.

**2.5 Optical Telescope**

The optical aspect telescope comprises a spherical mirror with an aperture 12 cm in diameter and a focal length of 2 m, which focuses stellar images onto a CCD camera. The CCD camera is a slightly modified version of the cameras developed by LLNL for the Clementine lunar mapping mission. The camera, which incorporates its own circuits for reading out image data through a 12-channel parallel interface and for processing camera configuration commands, is extremely compact (89 x 73 x 24 mm) and light (250gm). It employs a Thomson Th7867 frame-transfer CCD having array dimensions of 288 x 384 pixels, the pixel dimension being 23μm (~1 arcsec). The CCD chip has a phosphor coating, which extends the wavelength response down to 1800Å. The chip is cooled to -20 C, in order to minimize thermal noise on the output signal. The integration time for one CCD image is 0.906 s, determined by the need to obtain adequate signal-to-noise when observing the flight target, the white dwarf star G191-B2B (V=11.7) and surrounding stars in the 11 x 15 arc min telescope field.

**2.6 Spectrometer calibration**

The complete spectrometer was tested using an NRL test facility. The primary purpose of this work was to align and focus the optics of the assembled J-PEX spectrometer payload and no monitor counter was available for measurement of the effective area. The payload was mounted on an air-supported table, which provided isolation from building disturbances, and whose surface was an optical bench equipped for mounting optics and light sources. Final adjustment of the focus at EUV wavelengths was performed after attaching the payload to a vacuum chamber, which was equipped with a Penning discharge EUV light source and an off-axis paraboidal mirror producing a parallel beam. This beam of predominantly line radiation illuminated the spectrometer aperture. Grating adjustments were performed using temporarily installed motor-driven microactuators. The payload and EUV test chamber were pumped by the ground support equipment (GSE), specially designed for pumping the payload during integration and on the launch pad, and a large turbopump.

Fig. 3 shows typical spectra of He II lines produced by the Penning source, using pure He, equipped with a pinhole aperture 50μm in diameter. The line profiles were analyzed and a full-width at half maximum (FWHM) derived, in order to judge the quality of focus. Ray-tracing calculations predict that the grating image FWHM at the central wavelength (235 Å) should be 23μm. Combined in an RMS sum with the detector resolution of 20μm (FWHM, section 3.3) this predicts a width of 30.5μm (0.041 Å FWHM or R = 5754). Assuming an ACS pointing uncertainty of 20μm (~1arc sec, section 3.5) we would expect a spectral line width of 38μm or 0.051Å (R = 4618) in flight. However, the line width of 30.5μm (0.041 Å, R = 5754), should be compared with the measured values shown in Table 2, which are lower by 40-52 percent.

| Grating | Line width after deconvolution of pinhole diameter (μm FWHM) | Line width (Å FWHM) | Resolving Power |
|---|---|---|---|
| G3 | 63 | 0.087 | 2790 |
| G4 | 61 | 0.085 | 2870 |
| G1 | 59 | 0.082 | 2980 |
| G6 | 75 | 0.104 | 2330 |

Table 2: Measured widths of the 243 A He II calibration lines produced by the 4 J-PEX spectrometer gratings flown on NASA 36.195, and the corresponding spectral resolving power.

We attribute this difference between measurement and prediction to thermally induced image drift, distortion of the grating by the mounting structure, and partial illumination of the gratings. The limited beam intensity of the test facility required integrations of ~2 hr to obtain satisfactory images. The heat dissipated by the detector electronics caused its temperature to rise slowly during this time, reaching values above 30° C. The HeII line images drifted up to 60μm, comparable to the expected line width. As no movement of low-level MCP hotspots was measured, we conclude that thermal expansion was the cause. However, the measured performance represents a lower limit to that expected during flight, since the thermal timescale is very much longer than the ~10 minute duration (6 minutes with instruments switched on) of a rocket flight.

Although the thermal effects were the dominant factor degrading the measured spectral resolution, two additional effects make secondary contributions. Image distortions were also detected on occasion after a grating had been locked in position. In addition, the design of the Penning discharge EUV light source, in which the plasma discharge sits 10cm behind a pinhole aperture, does not fully illuminate the 40cm diameter collimating mirror and hence results in a partial illumination of each grating, such that peak flux is attained near one corner.

## 3. J-PEX FLIGHT ON NASA ROCKET 36.195DG

**3.1 Science goals**

The hot DA white dwarf G191-B2B is one of the brightest and best studied of the H-rich stars. However, a complete understanding of the EUVE spectrum of G191-B2B has been quite elusive. Lying near the top of the DA white dwarf cooling sequence, measurements of its effective temperature, surface gravity and composition represent an important benchmark in the study of the whole DA sample. From an assemblage of high spectral resolution observations with IUE and HST, it has been clear for some time that G191-B2B falls into the group of very hot DA stars with temperatures in excess of 50000K that contain significant quantities of heavy elements in their atmospheres. In particular detections of C, N, O, Si, S, P, Fe and Ni have all been reported in various papers[8,9,10,11,12]. Such material is responsible for a severe depression of the EUV flux in G191-B2B, when compared to the stars with pure H atmospheres. Indeed, this flux deficit as been used to identify similar objects in the EUV sky surveys of the ROSAT Wide Field Camera and Extreme Ultraviolet Explorer (EUVE). G191-B2B has been an important target for spectroscopic observations in the EUV to determine which sources of opacity are primarily responsible for the observed effect. In addition, an important goal has been to obtain a self-consistent model for the star, with an effective temperature, surface gravity and composition that can match the optical far-UV and EUV observations simultaneously. This would demonstrate our understanding of the nature of the star and the reliability of the model calculations, which can then be applied to other objects.

Initial attempts to match the observation with synthetic spectra failed to reproduce either the flux level or even the general shape of the continuum[13]. This problem was perceived to be caused by the lack of a sufficient number of Fe and Ni lines in the model. Adding some 9 million predicted lines to the few thousand with measured wavelengths did provide a self-consistent model able to reproduce the EUV, UV and optical spectra[14]. However, good agreement between the observed EUVE spectrum and the model prediction could only be achieved by inclusion of significant quantity of helium, either in the stellar photosphere or an ionized interstellar/circumstellar component. Unfortunately, due to the limited resolution of EUVE (~0.5Å) in the region of the HeII Lyman series, this inferred He contribution could not be directly detected.

More recently, it has been shown that photospheric heavy elements may not be distributed homogeneously (by depth) and that more complex stratified structures must be considered if the overall spectral distribution across the soft X-ray, EUV and far-UV bands is to be reconciled with the models[2,15]. Important progress has also been made in incorporating radiative levitation and diffusion self-consistently into the atmosphere calculations[15,16]. The need for a He contribution is reduced in these stratified models but not eliminated.

The capabilities of the J-PEX spectrometer proved a first opportunity to resolve some unanswered questions for G191-B2B. The first successful flight of J-PEX, on-board NASA Black Brant IX rocket 36.195DG, took place at 05.45UT on 2001 February 22nd. The payload completed its mission flawlessly and obtained a 300s exposure on the target. We present here the results of a search for the ionized HeII component predicted to be present in the EUV spectrum.

### 3. 2 Data reduction

The stellar spectrum appears as four offset lines across the detector image (Fig. 4), each covering a similar wavelength range. The image needs to be corrected for pointing drifts before the spectra can be extracted from it. A wavelength calibration is then applied, before all four spectra are co-added to yield the maximum signal-to-noise. We describe the process briefly below.

Aspect reconstruction was achieved by determining the centroid position of the EUV mirror image as a function of time, using the time tag attached to each MCP event. The corrected image was then rotated to align the spectra with the coordinate axes (Fig. 5). Once the dispersion axis of each spectral track is parallel to the horizontal coordinate axis, the x location of each event is related to wavelength. A suite of IDL routines was used to extract the accumulated photon event in each pixel, over sampling the spectral resolution by a factor 5 (bin size ~0.01Å) to avoid any loss of information.

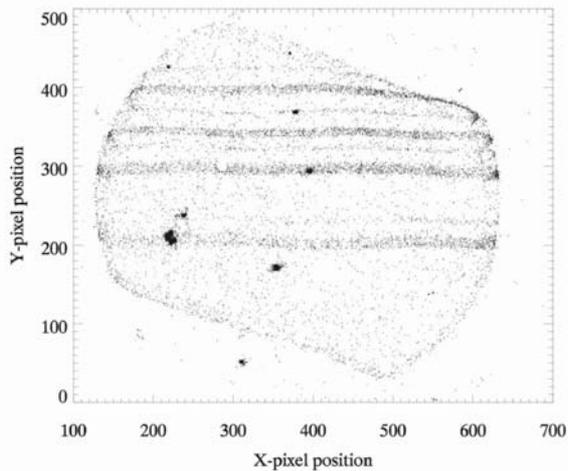
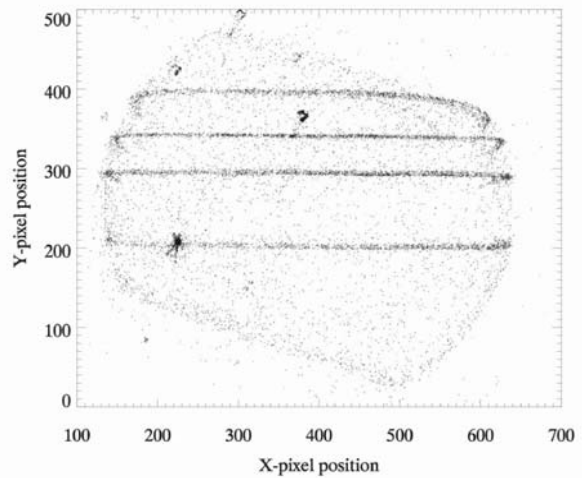

Fig. 4: J-PEX MCP detector image integrated for the full flight exposure, showing the 4 grating tracks. These are spread out due to in-flight motion of the pointing axis. The bright spot near the left edge of the lower spectrum is the direct image of G191-B2B.

Fig. 5: J-PEX MCP detector image integrated for the full flight exposure, showing the 4 grating tracks after correction for the attitude drift using the optical and EUV images.

Using laboratory calibration images of the Penning discharge source, an approximate wavelength scale was established for each spectrum, described by four separate 3rd order polynomials. Events were then summed in each bin to produce the four "raw" spectra, which had common bin sizes, but slight offsets in the central wavelengths due to small to the resolution of EUVE (~0.5Å) and compared with the G191-B2B medium wavelength spectrum, obtained with that instrument. This process also allowed us to determine the coarse wavelength scale offset for each grating. Fine offsets were determined by cross-correlating the individual spectra with each other to find common spectral features. Finally, the raw spectra were corrected for all offsets and re-binned to a common 0.024Å grid for co-addition. The final spectrum (Fig. 6) was rebinned to 0.048Å to optimize the signal-to-noise (5.0) for subsequent analysis.

Pre-flight dark exposures revealed a detector noise count rate of 4.2 counts s$^{-1}$ over the entire imaging area. Hence, the mean count per spectral bin accumulated during the exposure is sufficiently low that the stellar spectra could be assume to be free of background, apart from a single hotspot on the detector coinciding with one spectrum. This was removed by subtracting an estimate of the hotspot count rate, determined from adjacent pixels, from the spectrum. In addition, the EUV mirror image cut across one other spectrum. As this was too bright to subtract sensibly, the affected pixels were set to zero and an exposure correction applied to normalize that region when the spectra were co-added. Count errors were assigned on the basis of Poissonian statistics. The spectrum contains one flaw caused by the pointing on target, in which data above ~239Å was lost in two of the spectra. Thus only half the instrument effective area was utilized in this region, and accordingly the bin width was increased to 0.096Å.

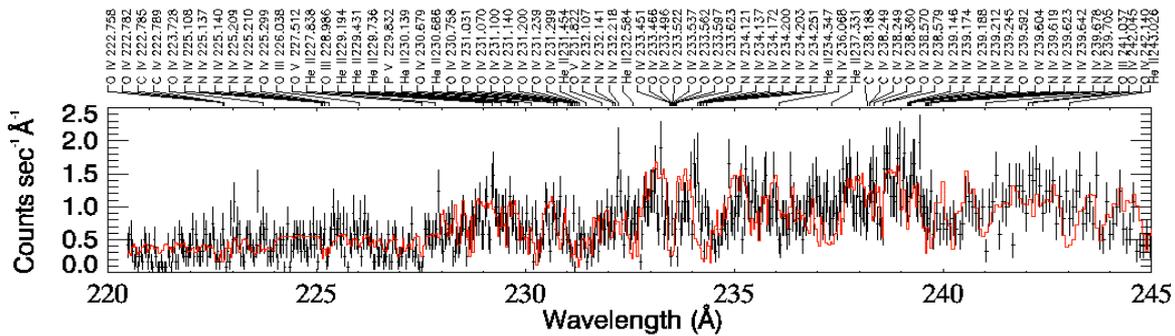

Fig. 6: High-resolution spectrum of G191-B2B obtained by the J-PEX spectrometer in the band 221-244 Å (error bars). The red histogram is the best-fit model of the star and ISM (see text). The strongest predicted lines of He, C, N, O, and P are labeled by ionization state and wavelength. Fe and Ni lines, too numerous to list, cause the unlabelled features and broader absorption structures.

### 3.3 Science Analysis

We have compared the observed spectrum (error bars in figure 6) with predictions based on a homogeneous composition stellar atmosphere and including interstellar HI, HeI and HeII absorption[17]. Although it has been established that stratified models give better overall description of the atmosphere of G191-B2B, the homogeneous models are perfectly adequate over narrow wavelength regions such as this and give a useful baseline for comparison with abundance measurements made in other studies. The analysis technique used to compare models and data has been described extensively in several earlier papers[e.g.14]. Hence, we just give a brief overview here. We utilise the programme XSPEC to fold model spectra through the J-PEX instrument response, taking into account the instrumental effective area and nominal spectral resolution. As we are dealing with a spectrum where there are small number of counts per bin, the best match between model and data was obtained by minimization of a Cash statistic[18]. This statistic does not assign an absolute value to the goodness of fit but does allow uncertainty ranges to be determined for each free parameter.

The spectral models were calculated using the non-LTE code TLUSTY[19]. These are based on work reported by Lanz *et al*[12] and Barstow, Hubeny & Holberg[1,20]. For this analysis, we fixed the stellar temperature and surface gravity ($T_{eff}$=54000K, log g=7.5) at the grid points closest to the values determine from the Balmer and Lyman lines[16]. Apart from the helium abundance, which was allowed to vary freely between the grid limits of $10^{-4}$ and $10^{-6}$, the abundances of the heavy elements were fixed at the values determined from earlier homogeneous analyses of G191-B2B (C/H=$4.0\times10^{-7}$, N/H=$1.6\times10^{-7}$, O/H=$9.6\times10^{-7}$, Si/H=$3.0\times10^{-7}$, Fe/H=$1.0\times10^{-5}$, Ni/H=$5.0\times10^{-7}$), but taking into account that the CIV lines near 1550Å have subsequently been resolved into multiple components by STIS[21].

The value taken for Fe/H lies between limits established by both FUV[22] ($2.4\times10^{-6}$) and EUV[2] ($3-4\times10^{-5}$) analyses. The effect of Fe/H in the 225-245Å is to change the level of the overall spectrum and we have verified that this uncertainty does not affect the conclusions reached in our analysis. The interstellar HI and HeI column densities were fixed at values obtained from analysis of the broader band, lower resolution EUVE spectrum[2] (HI=$2.15\times10^{18}$cm$^{-2}$, HeI=$2.18\times10^{17}$cm$^{-2}$). The parameters varied during the fit were the column density of HeII in the line of sight ($N_{HeII}$) and the photospheric He abundance (He/H by number). For this fit, the data were summed in bins of 0.06Å, equivalent to R=4000. Given the uncertainty in the resolving power (R=3000-4000), the fits were performed also for lower values of R, but yielded no significant change in the results presented here. The best fit to the data, shown by the grey line in Fig. 6, was obtained for $N_{HeII}$=$5.97\times10^{17}$cm$^{-2}$ and He/H=$1.6\times10^{-5}$. In Fig. 7 we show the 1, 2, 3, 5, and 10σ contours for the two parameters. The 3σ contour is the locus on which $\chi^2$ exceeds the minimum by 11.8 and within which the parameter confidence level is greater than 99.7%. We use this contour to derive 99% confidence limits of (5.76-6.18)$\times10^{17}$cm$^{-2}$ for $N_{HeII}$ and (1.31-1.91)$\times10^{-5}$ for He/H.

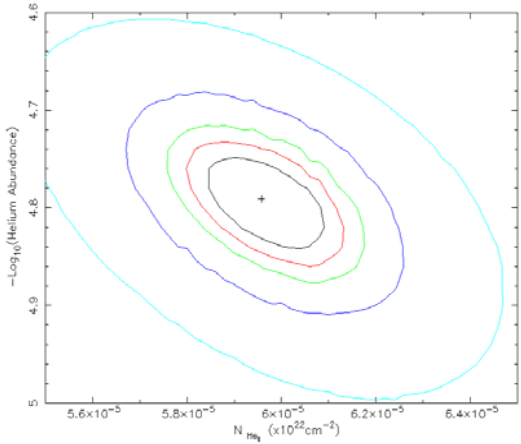

Fig. 7: The 1, 2, 3, 5, and 10σ contours for the two parameters varied in fitting a uniform-abundance white dwarf atmosphere model to the observed spectrum of G191-B2B. The 3σ contour corresponds to a confidence level of 99.7%.

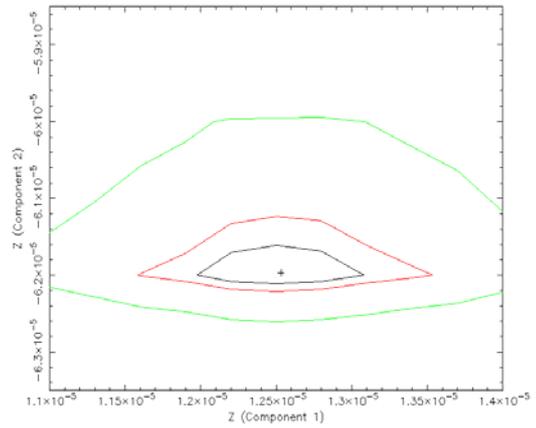

Fig. 8: The 3, 5, and 10σ contours for the relative redshifts when the interstellar HeII is assumed to be divided between two components. The 3σ contour corresponds to a confidence level of 99.7%.

The initial spectral analysis made the assumption that interstellar HeII present should be restricted to a single component. However, it has been revealed by FUV spectroscopy that there are two components in the ISM. One of these is associated with the local interstellar cloud (LIC), while the second is designated component I. They are separated in velocity space by 10km s$^{-1}$, equivalent to a relative redshift of $3.3\times10^{-5}$. Expanding on our first analysis, we have treated the observed interstellar HeII as two components with independent velocity and column density. Allowing these parameters to float freely produced a significantly better fit (by very much more than 10σ) compared to that assuming a single HeII component. Fig. 8 shows the 3, 5, and 10σ contours for the relative redshifts of each component, showing that a solution with equal redshift for each component is strongly excluded. For the best fit, the redshift difference between the two components is $7.5\times10^{-5}$, corresponding to a velocity separation of 22km s$^{-1}$. This is approximately twice the value expected from the FUV results. A finite HeII column density is assigned to both components, with values of $1.6\times10^{17}$cm$^{-2}$ and $3.5\times10^{17}$cm$^{-2}$ tentatively identified with the LIC and component I, respectively. The total column density of $5.1\times10^{17}$cm$^{-2}$ is 0.85 times that obtained for the best single component fit.

## 4. DISCUSSION

The good agreement between the best-fit model and the data in Fig. 6 is striking, e.g., at the prominent absorption feature at 233.5Å produced by a cluster of OIV lines. Many other features present are mainly blend of large numbers of FeV and NiV lines. The broad features between 227 and 232Å are characteristic of the overlapping series of interstellar HeII absorption lines superimposed on a stellar continuum. This is shown more clearly in Fig. 9, which shows the best fit with (top) and without (bottom) inclusion of interstellar HeII. The detailed shape of the upper spectrum coupled with the strong depression of the flux below 228Å is conclusive proof of the presence of interstellar HeII along the line of sight. Furthermore, for the upper panel, the He/H abundance ($1.6 \times 10^{-5}$) is consistent with the STIS limit ($2 \times 10^{-5}$), while in the lower panel the value is approximately four times greater ($8 \times 10^{-5}$).

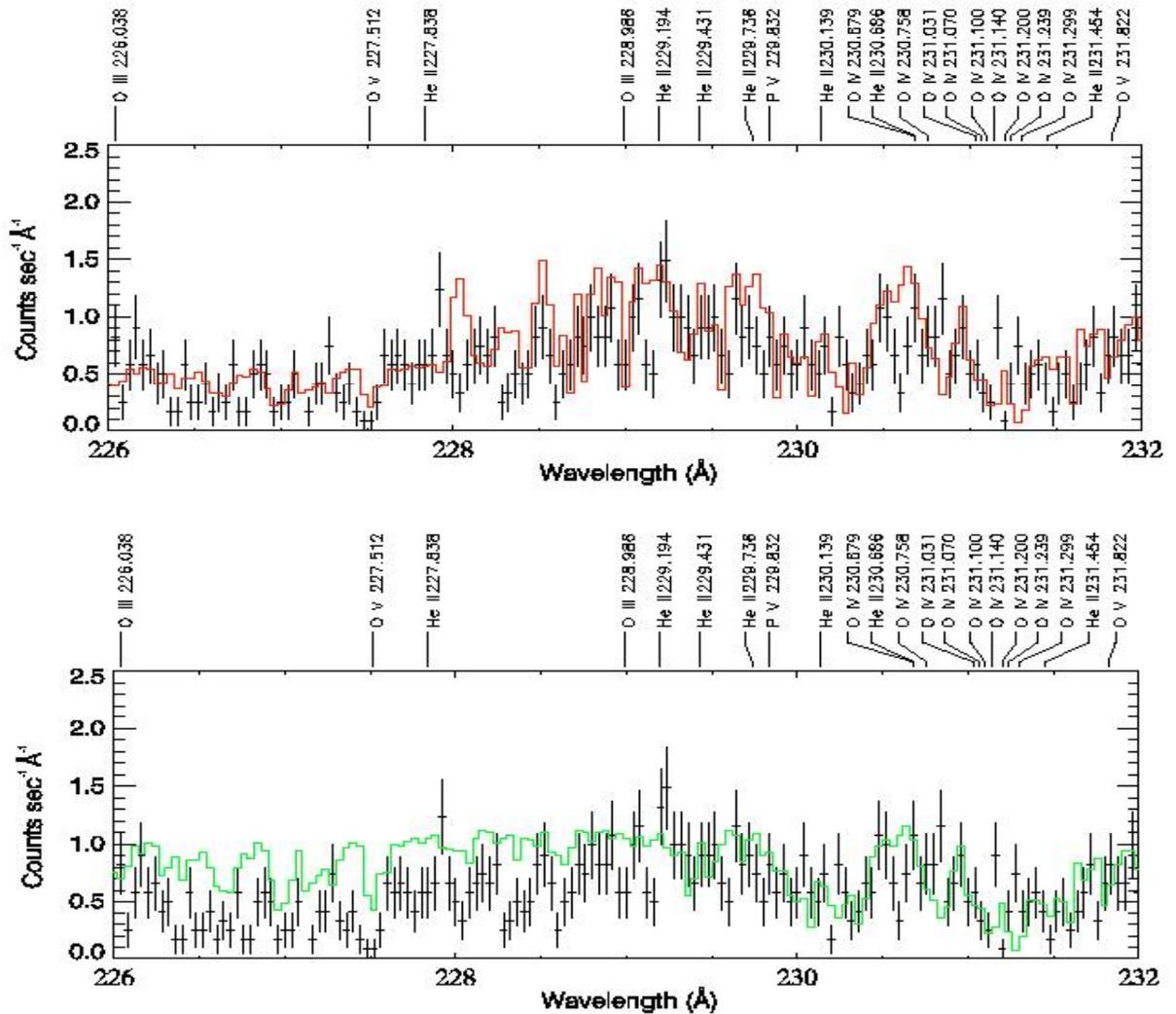

Fig. 9: Top – Expanded view of the J-PEX spectrum of G191-B2B in the wavelength range 226-232Å spanning the HeII Lyman series limit. The grey histogram is the best-fitting model of the stellar atmosphere and the ISM absorption. Bottom – Expanded view of the G191-B2B spectrum but this time showing the best fit that can be achieved without an interstellar or circumstellar HeII component

For the single component ISM fit, the measured column density of $5.97 \times 10^{17} \text{cm}^{-2}$ implies a He ionization fraction, based on EUVE measurements of the HeI column density, of ~0.73. This is much higher than the typical range of 0.25-0.50 in the local ISM[23]. However, if we assume that the HeII is split between the two interstellar components identified in the FUV, the ionization fraction may be lower. Since no HeI ISM features lie within the J-PEX wavelength range, it is not possible to assign a HeI column to either ISM component. However, the presence of the CIV in the FUV indicates that component I is highly ionized. Consequently, an estimate of the He ionization in the LIC can be obtained by assuming that it contains all the detected HeI, giving a value of 42% for HeII/He, although formally, this should be considered to represent a lower limit. This picture of two absorbing components along the line-of-sight is more consistent with other observations of the interstellar He ionization fraction. However, the origin of the more highly ionized material is still a mystery.

The J-PEX instrument is the highest resolution spectrometer yet developed for EUV astronomy. This presents special problems for measuring the performance using existing calibration facilities. For example, the spectral resolution recorded on the ground (R~2300-2900) was approximately half the target resolution of R=5000. Furthermore, it was not possible to carry out an end-to-end measurement of the spectrometer effective area, which has been calculated from measurements of the individual components. Consequently, observation of a well-studied source for which the flux an element composition are largely known can potentially provide important calibration information and confirmation of the anticipated instrument performance.

The spectral analysis described above, assumed a resolving power of 4000. The resulting model line profiles are a good match to the data, but, at the signal-to-noise achieved in the 300s exposure, it is not possible to distinguish between R=3000, 4000 or 5000, for any single absorption line. However, it is clear that the performance of the spectrometer is not worse than the ground calibration. The improved two-component ISM fit to the data provides a strong clue to the real performance of the spectrometer, since it deals with the combined effect of a larger number of lines. At first glance, the apparent velocity separation of 22km s$^{-1}$, which corresponds to a $\delta\lambda$~0.02Å, is beyond that achievable with the nominal 5000 resolving power ($\delta\lambda$=0.06Å). However, this is an overly simplistic view of how absorption lines may be separated, a process which depends on the intrinsic width of the lines and precise shape of the line response function. A good example is the separation of the G191-B2B CIV doublet into two components in the STIS spectrum. The wavelength separation is nominally 0.08Å (15km s$^{-1}$), compared to a resolution of 0.04Å, but the photospheric line is about twice the width of the interstellar/circumstellar component, producing an asymmetric blend rather than a truly split feature. In the J-PEX case, we are separating low volume density interstellar components, which produce intrinsically narrow lines. Hence, the observed line widths will be dominated by the instrument response. In this spectrometer design the line response function is far from gaussian, with a narrow peak superimposed on broad wings. Therefore, it is not unrealistic to distinguish the interstellar components at this level. It must be remembered that the division into two ISM components is not directly visible in the spectrum, at this signal-to-noise, and is just implied from the statistical improvement in the fit. Furthermore, a true test of the ability of the spectrometer to distinguish such components at this small velocity separation will required a detailed simulation with the full line response function. We will address this issue in a future paper. Nevertheless, the results presented here are a strong indication that the spectrometer is operating close to its nominal spectral performance.

## 5. CONCLUSIONS

We have presented an analysis of the first high-resolution spectrum of G191-B2B, obtained with the J-PEX spectrometer, which has the highest resolving power (3000-4000) achieved so far in the EUV and X-ray wavebands. The results show conclusively that ionized interstellar He is present along the line of sight to the star. Further analysis, treating this HeII as two components produces an improved fit, dividing the material between the LIC and component I. If confirmed, this would bring the inferred LIC He ionization fraction into line with other measurements. However, the source of the more highly ionized absorbing component remains a mystery.

The white dwarf model that best fits the data includes a significant abundance of photospheric He (He/H=$1.6 \times 10^{-5}$). At the 99% confidence level, we can exclude models containing no photospheric He. However, the predicted strength of the strongest expected photospheric line at 243Å is close to the noise level. Hence, we cannot claim a direct detection of He as yet.

The quality of the spectrum obtained during the J-PEX rocket flight underpins the ground-based measurements of the instrument performance. It will be necessary to carry out detailed simulations to obtain an objective assessment of the true performance, but the results do indicate that the spectrometer was operating close to nominal and that no in-flight degradation occurred.

## ACKNOWLEDGEMENTS

The University of Leicester and the Mullard Space Science Laboratory acknowledge the support they received for this project from the Particle Physics and Astronomy Research Council in the UK. The Naval Research Laboratory (NRL) was supported by NASA in this work, under grant NDPR S-47440F, and by the Office of Naval Research under NRL work unit 3641 (Application of Multilayer Coated Optics to Remote Sensing).

* mab@star.le.ac.uk; phone +44 116 252 3492; fax +44 116 252 3311; http://www.star.le.ac.uk/~mab; Department of Physics and Astronomy, University of Leicester, University Road, Leicester, LE1 7RH, UK.